# Easily generating and absorbing waves using machine learning

Yulin Xie[1], Xizeng Zhao[1,2,]*

*1. Ocean College, Zhejiang University, Zhoushan 316021, Zhejiang, China*

*2. The Engineering Research Center of Oceanic Sensing Technology and Equipment, Ministry of Education, China*

**\*** Corresponding author. E-mail: xizengzhao@zju.edu.cn

**Abstract:** High-order wave-making theories are becoming available but are limited to certain ranges of waves and wavemaker types in their applicability. Alternatively, machine learning can be considered to find nonlinear functional relationships. Therefore, this paper proposes a simple and universal framework for generating and absorbing waves based on machine learning. This framework trains neural networks to establish the transfer function between the free-surface elevation on the wavemaker and the wavemaker velocity. Significantly, penalty term and data augmentation techniques based on wave-making mechanisms are introduced to increase the generalization ability of neural networks, rather than pure data-driven. Therefore, once the target wave profiles in front of the wavemaker are given, it can realize generating waves and absorbing reflected waves at the same time. Taking piston and plunger wavemakers as examples, an in-house numerical solver is applied to simulate both wave generation and absorption. The simulated wave profiles and wave orbital velocities are validated with analytical solutions, showing that the proposed framework is effective at eliminating the re-reflection wave. Then, the validation for generating the solitary wave and the New-year wave is performed, indicating that the generated waves agree quite well with the desired wave elevation. The proposed framework can facilitate the wavemaker design in the future, and no complex theoretical derivation is required.

## 1. Introduction

Hydraulic flumes are usually equipped with a wavemaker at one end to generate two-dimensional wave trains. They generate waves by relating the wavemaker stroke to wave elevation with linear transfer functions, but the initial linear waves need a certain distance to evolve to the target nonlinear waves. Meanwhile, to reduce the reflections from the end wall of the flume, a common technique is to arrange a dissipative beach with a constant slope in front of the end wall. However, it will increase the length of the non-test domain of the flume. The dissipative beach cannot eliminate any waves reflecting back to the wavemaker if waves interacting with coastal or offshore structures are studied. Therefore, it is necessary to use the active wave absorption system. For active absorbing waves, early works employed digital filters or the use of linear transfer functions relating the wavemaker stroke to wave elevation (Schäffer and Klopman, 2000). However, the implementation of these methods is complex and difficult to be universal for wavemakers with arbitrary shapes, as well as unable to take the advantage of data fusion techniques. To this end, this work develops a universal method for active wave generation and absorption based on machine learning. This method aims to directly train neural networks to represent transfer functions mapping the desired surface-elevation at a front-mounted wave gauge to the wavemaker motion required to create it, which is applicable to wavemakers with arbitrary shapes.

For wave generation with absorbing wavemakers, the wavemaker movement is controlled to generate the desired incident waves and absorb reflected waves at the same time. With a time-domain or frequency-domain filter, the control signal can be obtained through a transformation of the wave signal in time. Hirakuchi et al. (1990)

developed a piston-type absorbing wavemaker with the water surface elevation at the mean position of the paddle as hydraulic feedback. Based on the force-feedback control, Spinneken and Swan (2009a) presented a mathematical model for an absorbing wavemaker, as well as an experimental verification. With free-surface elevation and orbital velocities at a fixed position in the fluid domain, Christensen and Frigaard (1994) proposed an active absorption system by using digital filters to estimate the absorption transfer function. Schäffer and Jakobsen (2003) have also developed a method to fit the frequency response of a digital filter to approximate the absorption transfer function by the linear theory, but they did not discuss the fitting technique. Based on the first-order wavemaker theory, Yang et al. (2016) utilized the iterative reweighted least-squares algorithm to approximate the absorption transfer function with an infinite impulse response digital filter. Although the above literature review introduces some previous works about active absorbing wavemakers, most researchers focused on the active absorption system in the frequency domain. In time-domain active absorption techniques, Schaffer and Klopman (2000) proposed a simple active wave absorption system according to the free-surface elevation measured at the wavemaker. The general assumption was linear long-crested wave theory in shallow water.

However, for the plunger-type wavemakers, the analytical solutions for wave generation are still difficult to obtain. Due to the shape of the plunger-type wavemaker more closely matching the velocity profile of near-surface water waves, the efficiency of the plunger is higher than that of the piston in deep water (Timmerberg et al., 2015). Wu (1988) used the boundary element method (BEM) to establish the transfer function for relating the stroke amplitude of plunger-type wavemaker to the far-field wave amplitude as a function of the wavenumber. More recently, He et al. (2021) proposed a theoretical method for the solitary wave generation with plungers and specified its constraints on the produced wave height. However, once the shape of the plunger is changed or an unusual shape is given, the formulae must be deduced again or even has no solution, especially when considering the gap (Wu, 1991; Nikseresht et al., 2020). Hence, further improvements are needed to develop control signals for the plunger-type devices with optimization procedures or nonlinear transfer functions (Hicks et al., 2021). The wave-making system proposed in this paper can map nonlinear transfer functions very easily, which will extend the functionalities of existing plunger-type wavemakers.

To establish transfer functions relating the wavemaker velocity to desired free-surface elevation in front of the wavemaker, most researchers focused on linear theory methods (Schaffer and Klopman, 2000; Didier and Neves, 2012) or infinite impulse response digital filter methods (Yang et al., 2016; De Mello et al., 2013; Spinneken and Swan, 2012). However, nonlinear wave generations need high-order wave-making theory, there is limited evidence of analytical solutions encompassing all features relevant in the applied ocean, and coastal research exists (Eldrup et al., 2019; Khait et al., 2019). Recently, Mahjouri et al. (2020) proposed a simple and practical active control algorithm for piston-type wavemakers by some constant gains. Though the control system did not use any transfer function or filter in the feedback and feed-forward loops, the constant gains must be adjusted in each test where the water depth and working frequency change. The implementation of these methods mentioned above may be complex and difficult to be universal for arbitrary wavemakers. Therefore, this work pays attention to the machine learning methods that possess strong nonlinear mapping ability. In recent years, machine learning has attracted more and more attention in the field of fluid mechanics (Kutz et al., 2017; Chen et al., 2021; Lee et al., 2019). Artificial neural network (ANN) is one of the most widely used structures in machine learning methods, which is represented as a set of interconnected neurons. These neurons function in numerical form, and there are sequential multiplication accumulation operations connecting neurons. The application of neural networks to predict wavemaker inputs based on specific wave traces for the calibration of waves close to the breaking limit was implemented by Schmitt (2017). Their work indicates that neural networks are a valuable tool for calibrating wavemaker and can produce better results with further experiments. However, to the best of our knowledge, there is no systematic study about establishing a universal framework to map the absorption transfer function relating the wavemaker velocity to desired free-surface elevation in front of the wavemaker by the ANNs.

To that end, the central aim of this work is to introduce the transfer function mapped by the ANNs in our numerical wave flume, including the data generation, data augmentation technology and hyper-parameters sensitivity analysis during the training process. The work proceeds as follows. For model training, waves are first generated by the sinusoidal motion of the wavemaker. The monitored surface elevations were then used as model input and the given wavemaker velocity input was used as the desired model output. With running different sinusoidal motions of wavemaker, more data is available and the generalization ability of the model will become stronger. Then, the wavemaker velocity will be modified to match the velocity induced by the waves that need to be absorbed in practical application. The advantages of this wave absorption system are two-fold. Firstly, a wide variety of wave conditions with arbitrary shape wavemakers can be applied quickly. Secondly, the transfer function required to absorb reflecting waves can also be used to generate a specified wave.

The rest of the paper is organized as follows. The numerical wave flume used for performing the simulation is introduced in section 2, including the Navier-Stokes equation solver and level set method. In section 3, the wave-making methodology based on ANNs is outlined, including the general framework of wavemakers and the structure of ANNs. In section 4, the piston-type absorbing wavemaker is tested. The data generation and the influence of the hyper-parameters on the mapping accuracy are also introduced. In section 5, the wedge-shaped plunger absorbing wavemaker is tested for both regular and irregular waves. Especially, data augmentation is introduced to increase the mapping accuracy. In section 6, the cylinder-shaped plunger absorbing wavemaker is validated for generating solitary waves and New-year waves. The main conclusions are drawn in section 7.

## 2. Numerical method

In this section, the numerical method utilized for an in-house developed CFD solver is introduced, including the Navier-Stokes equation solver and level set method. Some validations of the numerical method can refer to our previous work (Xie et al., 2021a, Xie et al.,2021b).

### 2.1 Navier-Stokes equation solver

An in-house developed CFD solver is established based on the finite difference method. The incompressible and viscous Navier-Stokes equations are expressed as:

$$\nabla \cdot \mathbf{u} = 0, \tag{1}$$

$$\frac{\partial \mathbf{u}}{\partial t} + (\mathbf{u} \cdot \nabla)\mathbf{u} = -\frac{1}{\rho}\nabla p + \frac{\mu}{\rho}\nabla^2 \mathbf{u} + \mathbf{g}, \tag{2}$$

where $\mathbf{u}$ is the velocity vector, $p$ is the pressure, and $\mathbf{g}$ is the gravitational acceleration. $\rho$ and $\mu$ denote the density and viscosity, respectively.

The prediction-correction fractional steps solution scheme (Chorin, 1968) is utilized to solve the Navier-Stokes equations, and the constraint of mass conservation is achieved by coupling the pressure term with the continuity equation. Then, an intermediate velocity term is introduced to decouple the pressure term from the momentum equation. Based on the non-incremental pressure correction, the velocity field and pressure field of the fluid at the next moment can be obtained. The advection term in Eq. (2) is calculated by a third-order finite-difference scheme: Cubic Interpolated Pseudo particle (CIP) scheme (Yabe et al., 2001). The second part of the non-advection term in Eq. (2) is calculated by a central difference method. The pressure Poisson equation is solved by an algebraic multigrid solver. The details of the Navier-Stokes equation solver used in this paper can refer to our previous works (Zhao and Hu, 2012; Zhao et al., 2014). In this study, the ghost-node immersed boundary method (Peskin, 2002; Zhang and Zheng, 2007; Calderer et al., 2014) is used to deal with the fluid-solid interaction. The ghost nodes are imposed to present the effect of the moving body.

## 2.2 Level set method

As a front capturing technique for interfaces, the level set method was applied widely to calculate the free surface flows (Osher and Fedkiw, 2001; Osher, et al., 2004). The basic idea behind the level set method is forming a high-dimensional function to handle topological changes, and then the interface can be viewed as the zero level sets of the function. Under a given velocity field, the transport of the level set function can be regarded as solving the advection equation. In this study, the semi-Lagrangian scheme is used for solving the advection equation of the level set function, which possesses good stability (van Leer, 1979; Seaïd, 2002).

The level set function $\Phi(X,t)$ can be defined as a signed function, where the liquid phase is positive and the gas phase is negative. The interface $\Gamma$ can be expressed as:

$$\Gamma = \{X | \Phi(X,t) = 0\}, \tag{3}$$

$$\Phi(X, t=0) = \begin{cases} d(X, \Gamma) & \text{for } X \text{ in the liquid,} \\ 0 & \text{for } X \in \Gamma, \\ -d(X, \Gamma) & \text{for } X \text{ in the gas,} \end{cases} \tag{4}$$

where $d(X, \Gamma)$ is the distance function. The density and viscosity in each fluid can take on two different values depending on the sign of $\Phi(X,t)$, can be expressed as:

$$\rho(\bar{\Phi}) = \rho_G + (\rho_L - \rho_G)\bar{H}(\bar{\Phi}), \tag{5}$$

$$\mu(\bar{\Phi}) = \mu_G + (\mu_L - \mu_G)\bar{H}(\bar{\Phi}), \tag{6}$$

where the subscripts $G$ and $L$ represent the gas and liquid phases, respectively.

## 3. Wave making methodology based on ANNs

In this study, we directly establish the transfer function between the free-surface elevation in front of the wavemaker and the wavemaker velocity. Therefore, once the target wave profiles in front of the wavemaker are given, we can realize generating waves and absorbing reflected waves at the same time. The target waves can be generated near the wavemaker without a long-distance propagation process, which can shorten the length of the flume. The work reported in this paper uses the numerical wave flume for the preparation of suitable training data but the method should be applicable to any numerical and physical wavemaker.

### 3.1 General framework

According to Schmitt et al. (2021), the problem of finding the transfer function can be illustrated in the following way. Firstly, we have an output variable $Vp_i$, the required wavemaker velocity, defined as a function $F(\eta_i, V\eta_i…W_i)$ dependent on the hydraulic feedbacks including free-surface elevation in front of the wavemaker and some wave parameters. By forcing sinusoidal motion of the wavemaker, we can obtain a set of vectors $S_i$ which means the set of observations ($\eta_i, V\eta_i…W_i$) at the probe mounted in front of the wavemaker. The inputs and outputs can be defined by Eq. (7)

$$F(S(\eta, V\eta..., W)) \to Vp \tag{7}$$

Then, the universal approximation feature of neural networks supports us to find an approximation for the unknown function $F$ associating the variable $Vp_i$ with the vectors $S_i$ from a series of collected samples. Fig. 1 shows the block diagram of the proposed framework. In the main control loop, the measured free-surface elevation $\eta_\mathrm{m}$ includes the generated, reflected, and re-reflected waves. The desired free-surface elevation $\eta_\mathrm{d}$ is provided as a data file. Therefore, the free-surface elevation of the reflected waves, $\eta_\mathrm{R}$, is to be absorbed, by comparing the desired free-surface elevation $\eta_d$, with the measured one in front of the wavemaker, $\eta_m$. Then, the wavemaker velocity will consider the velocity induced by the wave that will be absorbed. The feedback signal consists of the measured free-surface elevation $\eta_m$, desired free-surface elevation $\eta_d$, wavemaker position and wave parameters. The wave elevation control can be regarded as equivalent to the velocity control of the wavemaker.

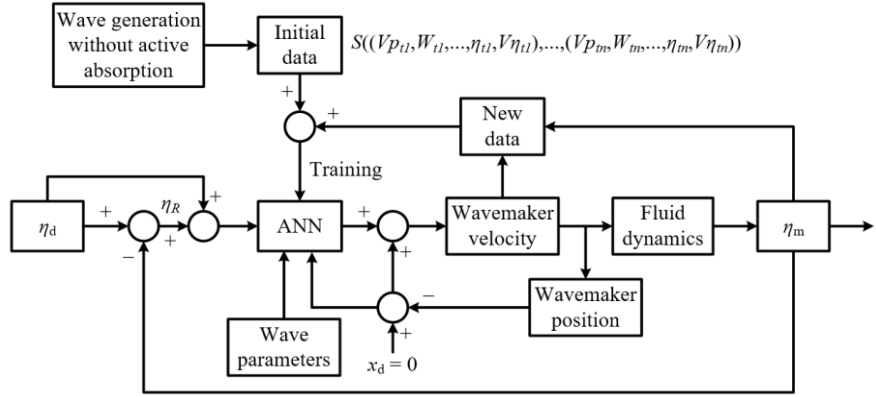

Fig. 1 Block diagram of the implemented absorbing wavemaker framework

However, it should be noted that controlling the wavemaker velocity does not necessarily guarantee the position control of the wavemaker. Possible drift in the position of the wavemaker may saturate the wavemaker stroke during the wave elevation control. To prevent this drift, closed-loop position control of the wavemaker is employed. This action needs to be slow and smooth enough to avoid any disturbance of the wave absorption loop. The initial zero-position '$x_d=0$' is used as the desired input and a smoother transition (e.g. a power function) is used to regulate the wavemaker.

**3.2 Structure of ANNs**

The multi-layer perceptron (MLP) model is utilized for the structure of ANNs. As long as the activation function is continuous, bounded and non-constant, the multi-layer feed-forward neural network can approximate any well-behaved function (Hornik, 1991). Once trained, any required free-surface elevation in front of the wavemaker can be generated without further iterative calibration steps. The detailed structure of the ANNs is illustrated in Fig. 2. The input variables mean the observations ($\eta_i$, $V\eta_i$…$W_i$). The left scatter plots of Fig. 2 present a series of collected samples. The input variables include surface elevation at the wave gauge and other monitor parameters. The output variable is the required wavemaker velocity. For the training of the ANNs, the initial process is forcing the sinusoidal motion of the wavemaker to generate a series of regular waves. Of course, forcing the random motion of the wavemaker to generate a series of irregular waves is also acceptable. The transfer function is then trained by relating the wavemaker velocity to the measured observations.

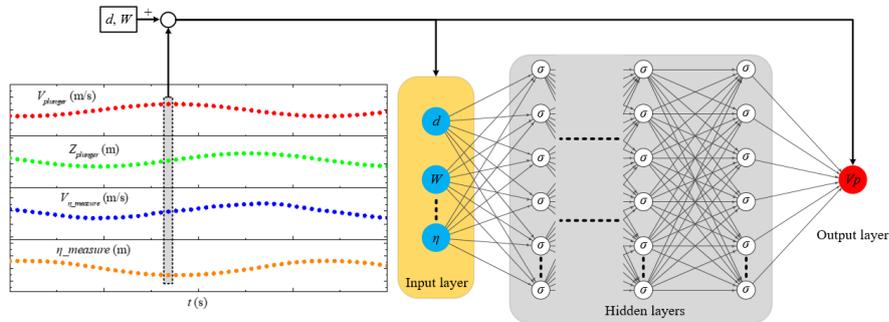

Fig. 2 Structure of the artificial neural network. The left scatter plots present an example of a sample set. The input variables include surface elevation at the wave gauge and other monitor parameters. The output variable is the required wavemaker velocity.

## 4. Case I: Piston-type absorbing wavemaker

The piston-type wavemaker is tested in this section, data generation for the training of the ANNs will be given in subsection 4.1. Then, the influence of the hyper-parameters on the mapping accuracy will be discussed in subsection 4.2. In subsection 4.3, regular wave generation with active wave absorption for full reflection case (i.e. a standing monochromatic wave) will be tested.

### 4.1 Data generation

In this part, to ensure that the training set is free from wave reflection interference, the wave is generated without active absorption but with a damping region at the end of the wave flume. The diagram of the 2D numerical wave flume is shown in Fig. 3. The still water depth is $d$ and the flume length is $12L$, where $L$ is the characteristic wavelength. The initial zero-position of the paddle is located in $x = 0$ m. Wave damping zones are placed on the right sides of the flume with a width of $6L$. The entire computational domain is discretized using a non-uniform grid distribution. The numerical time step $dt$ in this study is $1.0 \times 10^{-3}$ s. The wave gauge mounted in front of the wavemaker is employed to record the free-surface elevation.

To generate sufficient data for the training of the neural network models, multiple distinct instances of the input and output variables can be recorded by forcing the sinusoidal motion of the wavemaker to generate a series of regular waves. The cases for data generation can be arbitrary (forcing random motion of the wavemaker to generate a series of irregular waves is also acceptable), as long as its range can cover the period and wave height of the target wave. Here, we choose the sinusoidal motion of the wavemaker and the motion parameters are listed in Table 1. Each case only needs to be run about three periods of the sinusoidal motion of the wavemaker. It should be noted that the input variables defined in this part are just the free-surface elevation in front of the wavemaker, wave period (peak wave periods for irregular wave) and water depth.

Table 1 The sinusoidal motion parameters for data generation

|  |  | $T$=0.8 s | $T$=1.0 s | $T$=1.2 s | $T$=1.4 s |
|---|---|---|---|---|---|
|  | Stroke=0.04 m | Case1 | Case2 | Case3 | Case4 |
| $d$=0.4 m | Stroke=0.08 m | Case5 | Case6 | Case7 | Case8 |
|  | Stroke=0.12 m | Case9 | Case10 | Case11 | Case12 |
|  | Stroke=0.04 m | Case13 | Case14 | Case15 | Case16 |
| $d$=0.5 m | Stroke=0.08 m | Case17 | Case18 | Case19 | Case20 |
|  | Stroke=0.12 m | Case21 | Case22 | Case23 | Case24 |
|  | Stroke=0.04 m | Case25 | Case26 | Case27 | Case28 |
| $d$=0.6 m | Stroke=0.08 m | Case29 | Case30 | Case31 | Case32 |
|  | Stroke=0.12 m | Case33 | Case34 | Case35 | Case36 |

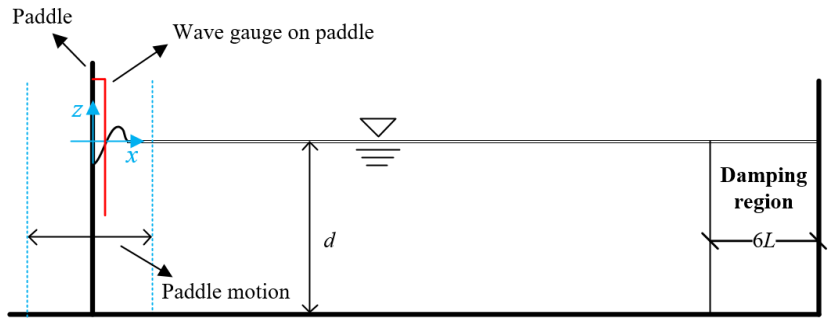

Fig. 3 Schematics of the 2D numerical wave flume

**4.2 Influence of the hyper-parameters on the mapping accuracy**

The hyper-parameters in an MLP model can affect the mapping accuracy. Here, we will discuss the number of layers used, the number of neurons per layer, the ratio of dropout layer (Srivastava et al., 2014) and the penalty term. The dropout layer is used to prevent neural networks from overfitting. The loss function used in the training process is defined as Eq. (8) where the second term of the left side is the penalty term. The penalty term is first proposed in this study to embed the physical feature into the MLP model and prevent neural networks from overfitting, which means that the wavemaker velocity and the free-surface elevation should be in the opposite direction.

$$\text{Loss} = \sum_{n=1}^{N} |Vp(S_n)_{\text{ANN}} - Vp_n|^2 + \lambda \sum_{n=1}^{N} |\text{sign}(Vp(S_n)_{\text{ANN}}) + \text{sign}(\eta_n)|^2 \qquad (8)$$

$$\text{MSE} = \sum_{n=1}^{N} |Vp(S_n)_{\text{ANN}} - Vp_n|^2 \qquad (9)$$

In this study, the activation function used at each layer is chosen as Rectified Linear (ReLU). The k-fold cross-validation is usually used to eliminate the bias in the training data. Therefore, the data samples will be divided in a 9 : 1 ratio of training to test samples and then the members of each set are chosen randomly. In this study, the mean of squares of errors (MSE) between the samples and the predictions during the training phase is used as the judgment criterion of mapping accuracy, shown in Eq. (9).

The basic values of these hyper-parameters are selected as that the number of layers is 4, the number of neurons per layer is 32, the ratio of dropout layer is 10% and the weight of the penalty term $\lambda$ is 1. The training epoch is fixed at 2400. Table 2 shows the test results of hidden layer parameters including the number of layers and the number of neurons per layer. As expected, we observe that as the number of layers and neurons is increased (e.g the capacity of the neural network to approximate more complex functions), the predictive accuracy is increased. The results showed that all tasks convergence with a MSE below 0.0001.

Table 2 MSE of the test for the hidden layer parameters

|          | Neurons =16 | Neurons =32 | Neurons =64 |
|----------|-------------|-------------|-------------|
| Layers=2 | 5.39038e-05 | 2.02104e-05 | 8.07028e-06 |
| Layers=4 | 4.13974e-05 | 1.75347e-05 | 6.85055e-06 |
| Layers=6 | 3.87137e-05 | 1.67195e-05 | 6.08309e-06 |

Table 3 MSE of the test for the overfitting prevention parameters

|                    | Dropout =0  | Dropout =10 | Dropout =20 |
|--------------------|-------------|-------------|-------------|
| Weight $\lambda$ =0 | 5.17951e-05 | 2.17839e-05 | 1.33114e-05 |
| Weight $\lambda$ =1 | 3.12597e-05 | 1.75347e-05 | 4.89697e-06 |
| Weight $\lambda$ =5 | 2.17839e-05 | 3.88151e-06 | 3.53083e-06 |

Table 3 shows the test results of the overfitting prevention parameters including the ratio of dropout layer and the weight of the penalty term $\lambda$. The results showed that as the ratio of the dropout layer and the weight of the penalty term are increased, the predictive accuracy is increased. The proposed penalty term in this study has played a positive role.

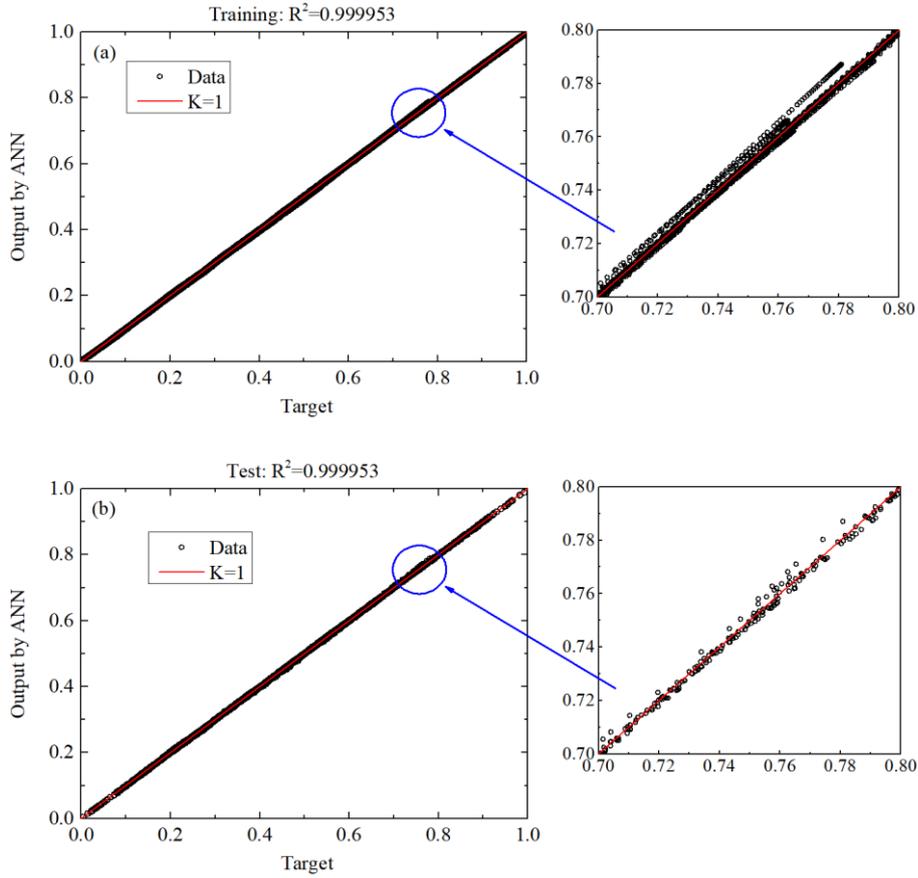

Fig. 4 The scatter plot of the estimated and the observed values of the normalized *Vp*. The correlation coefficient $R^2$ is listed at the top of each plot.

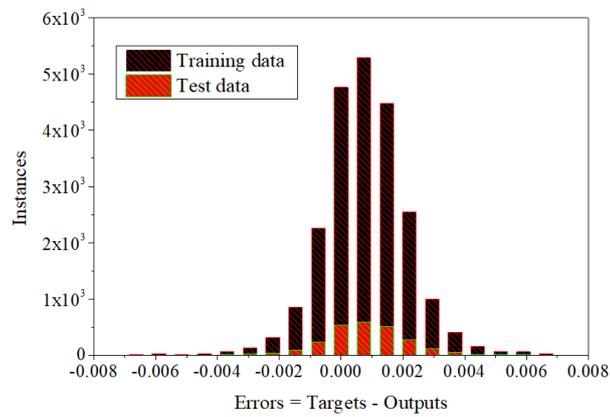

Fig. 5 The distribution of the error for the fit.

Based on the above analysis, we identified the parameters ultimately used for subsequent studies that the number of layers is 6, the number of neurons per layer is 64, the ratio of dropout layer is 20% and the weight of the penalty term $\lambda$ is 5. Fig. 4 shows the scatter plot of the estimated (outputs by ANN) and the observed (samples) values of the normalized *Vp*. If the fit were perfect, all the points should lie on the 45-degree line K. Fig. 4(a) shows the fit for the training data while Fig. 4(b) shows the same thing for the test data. The MPL model performs very well. Fig. 5 shows the error distribution for the final fit. Most of the training data points have close to zero error, but even the test points are closely distributed around the origin.

### 4.3 Validation of the learned ANN on absorbing wave-making

In this part, the performance of the proposed wave-making system in eliminating the spurious re-reflection of outgoing waves is demonstrated. Here, monochromatic wave generation with active wave absorption, a standing wave is expected to be generated in the computational domain. The diagram of the 2D numerical wave flume is shown in Fig. 6. The wavemaker is located at the one end while the wall is $8L$ away from the wavemaker. The targeting incident wave input to the wavemaker is a second-order Stokes wave. The incident wave period is 0.851 s and its amplitude is set as 0.018 m. Two wave gauges WG1 and WG2 are employed to record the wave evolution, located at antinode (equal to $L/2$ from the right-side wall) and node (equal to $L/4$ from the right-side wall), respectively.

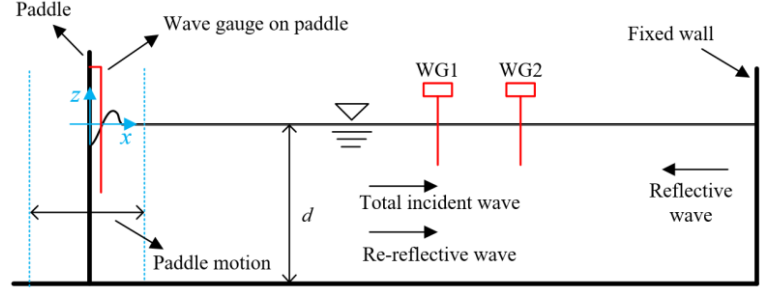

Fig. 6 Schematics of the 2D numerical wave flume

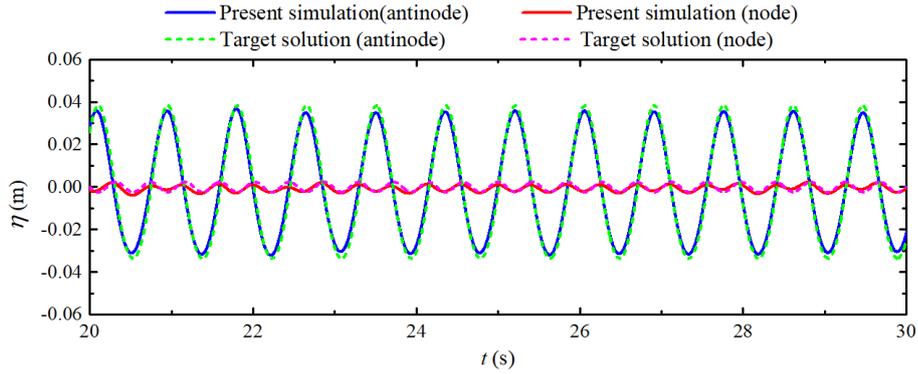

Fig. 7 Time series of the free-surface elevation at node and antinode.

Fig. 7 shows the time series of the free-surface elevation at node and antinode. The target solutions (i.e second-order Stokes solutions) are also plotted for comparison, assuming perfect reflection at the right-side wall. It is known that a perfect active wave absorption system can fully absorb the reflected wave energy and then prevent the incident wave from diverting. As shown in Fig. 7, the amplitudes at node are very small and the amplitudes at antinode are in good agreement with the expected second-order solution, proving the ability of the proposed wave-making system.

### 5. Case II: Wedge-shaped plunger absorbing wavemaker

For most real installations of plunger wavemakers, there is a finite gap between the wedge and the tank wall. Significant fluid motion can be induced in the gap, which may lead to a nonlinear phenomenon in the wave generated in front of the wedge (Nikseresht et al., 2020). The wave-making system proposed in this paper can consider the effect of the gap without extra effort, which will extend the functionalities of existing plunger-type wavemakers.

Therefore, both the wedge-shaped plungers with and without gap are studied in this section. Validation of wedge-induced wave and data generation will be given in subsection 5.1. Then, the influence of the data augmentation on the mapping accuracy will be discussed in subsection 5.2. In subsection 5.3, the trained ANNs will be applied to generate irregular waves by using the proposed framework illustrated in section 3. Then, regular wave generation with active wave absorption for full reflection case (i.e. a standing monochromatic wave) will be generated in subsection 5.4.

**5.1 Validation of wedge-induced wave and data generation**

To validate the numerical solver in the vertical movements of wedges and their induced impulse waves, both the wedge-induced waves with and without gap are compared with linear potential flow solutions obtained by Wu (1988) and the experiment results (Ellix and Arumugam, 1984). The diagram of the 2D numerical wave flume is shown in Fig. 8. The still water depth is $d = 0.5$ m and the flume length is $18L$, where $L$ is the characteristic wavelength. The initial (zero-position) bottom corner of plunger is located in $z = -D_{plunger} = -0.3$ m. The bottom angle of the wedge is $\theta$. The gap between the plunger and wall is 0.05 m. Wave damping zones are placed on the right sides of the flume with a width of $6L$. The entire computational domain is discretized using a non-uniform grid distribution. The numerical time step $dt$ in this study is $1.0 \times 10^{-3}$ s. A wave gauge mounted along the wavemaker is employed to record the free-surface elevation. The wave gauge WG1 located at $x = nL$ ($n = 6.5$) is employed to record the wave evolution.

Fig. 9 plots the $a/S_{plunger}$ ratio for different solutions versus the wavenumber, where $a$ is the record wave amplitude and $S_{plunger}$ is the stroke of the plunger. It is shown that the wedge-induced waves without gap simulated by the present numerical solver are in reasonable agreement with the theoretical solution in small $kh$, while those in reasonable agreement with the experiment data in large $kh$. However, the results of the wedge-induced waves with gap are consistent with the experiment data in almost all $kh$. These phenomena may be caused by energy dissipation, energy leakage around the gap of the plunger. Anyway, these results indicate the reliability of numerical solver in generating waves by a moving boundary.

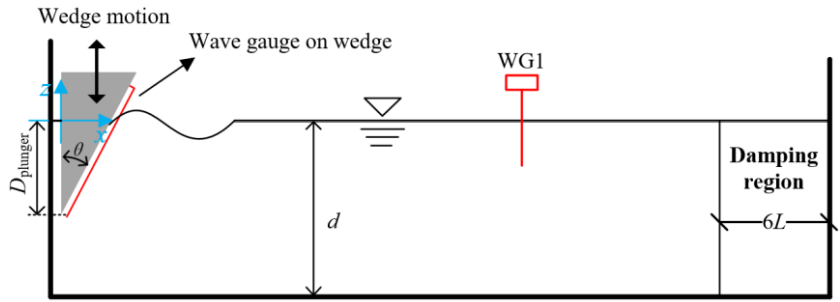

Fig. 8 Schematics of the 2D numerical wave flume

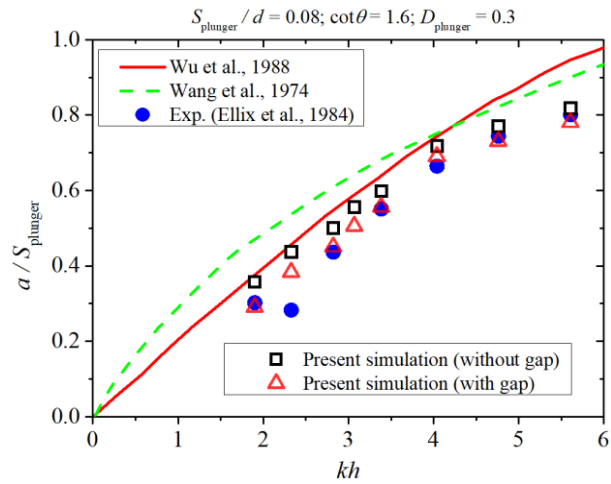

Fig. 9 Comparison of the wave amplitude to stroke amplitude ratio in the present work with experimental and theoretical transfer functions.

Similar to section 4.1, we force the sinusoidal motion of the wavemaker to generate a series of regular waves to generate sufficient data for the training of the neural network models. The diagram of the 2D numerical wave flume and the shape parameters of the wedge are consistent with those mentioned in the validation above. The motion parameters for wedge-induced waves without gap are listed in Table 4, while those for wedge-induced waves with gap are listed in Table 5. Each case only needs to be run about three periods of the sinusoidal motion of the wavemaker. It should be noted that the input variables defined in this part will be discussed deeply in the next subsection.

Table 4 The sinusoidal motion parameters of wedge-induced waves without gap for data generation

|  |  | $T$=0.8 s | $T$=1.0 s | $T$=1.2 s | $T$=1.4 s |
|---|---|---|---|---|---|
|  | Stroke=0.04 m | Case1 | Case2 | Case3 | Case4 |
| $d$=0.5 m | Stroke=0.08 m | Case5 | Case6 | Case7 | Case8 |
|  | Stroke=0.12 m | Case9 | Case10 | Case11 | Case12 |

Table 5 The sinusoidal motion parameters of wedge-induced waves with gap for data generation

|  |  | $T$=0.8 | $T$=0.9 | $T$=1.1 | $T$=1.4 |
|---|---|---|---|---|---|
|  | Stroke=0.06 m | Case1 | Case2 | Case3 | Case4 |
| $d$=0.5 m | Stroke=0.10 m | Case5 | Case6 | Case7 | Case8 |
|  | Stroke=0.14 m | Case9 | Case10 | Case11 | Case12 |

**5.2 Influence of the data augmentation on the mapping accuracy**

In this subsection, we will introduce data augmentation to improve the mapping accuracy of our MLP model. The hyper-parameters used in the MLP model can refer to that in section 4.2, which will not be discussed again. As mentioned above, the input variables defined in subsection 4.1 are just the free-surface elevation in front of the wavemaker, wave period (peak wave periods for irregular wave) and water depth. In this part, we discuss three input schemes from less to more variables, defined as follows.

Input scheme 1: wave period (peak wave periods for irregular wave); water depth; the free-surface elevation in front of the wavemaker.

Input scheme 2: wave period (peak wave periods for irregular wave); water depth; the free-surface elevation in front of the wavemaker; velocities of the free-surface elevation in front of the wavemaker.

Input scheme 3: wave period (peak wave periods for irregular wave); water depth; the free-surface elevation in front of the wavemaker; velocities of the free-surface elevation in front of the wavemaker; the position of wavemaker.

Fig. 10 shows the convergence curves for different input schemes during the training process, while Fig. 11 displays the distribution of the error for the fit of test data. Obviously, with the increase of input variables (i.e using data augmentation), the convergence error is smaller and the error distribution is more concentrated near zero error. From the perspective of neural networks, it can be explained that the richer the input information, the better the fitting accuracy. It should be noted that the velocities of the free-surface elevation in front of the wavemaker can be directly obtained from the recorded free-surface elevation by a numerical method such as first-order upwind. From the perspective of physical characteristics, velocities of the free-surface elevation in front of the wavemaker provide the momentum information, while the couple of free-surface elevation in front of the wavemaker and the position of the wavemaker provide the information of waterplane area (i.e mass information). It is known that momentum information and mass information are important for the derivation of wave-making theory. Hence, the proposed wavemaker system based on ANN is reasonable and interpretable.

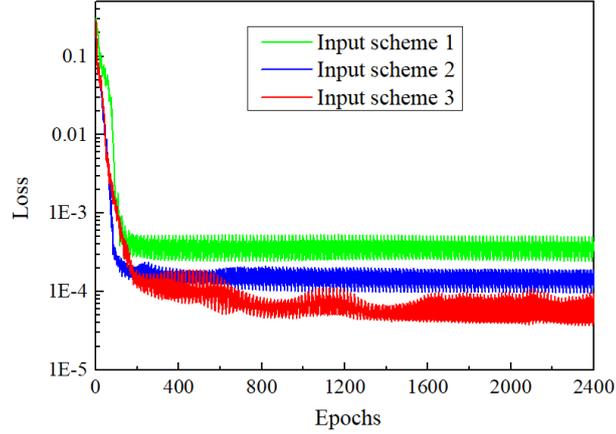

Fig. 10 Comparison of convergence curves for different input schemes during the training process.

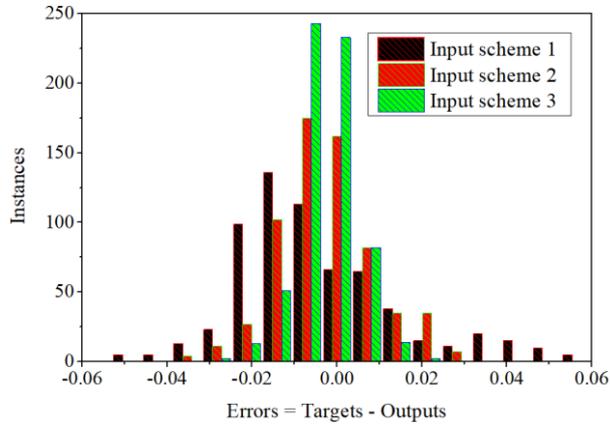

Fig. 11 The distribution of the error for the fit of test data.

## 5.3 Validation of the learned ANN on generating irregular wave

In this subsection, the performance of the proposed wave-making system is investigated through both irregular wave generation and absorption. The diagram of the 2D numerical wave flume is shown in Fig. 12. For assessing the performance of the system by the time series of free-surface elevation, the reflection wave is directly absorbed in the right side of the wave flume. Therefore, the wave is generated with the left side wavemaker while an active absorption wavemaker is set on the right side. The still water depth is 0.5 m and the flume length is $8L$, where $L$ is the characteristic wavelength. The shape parameters of the wedge are consistent with those mentioned in subsection 5.1, including the gap size. Two wave gauges WG1 and WG2 are employed to record the wave evolution, located at $x = nL$ ($n$ = 4, 4.5), respectively. The P-M spectrum is used as the target irregular wave spectrum. The irregular waves with peak wave periods $T_p$ = 1.3 s and significant wave heights $H_s$ = 0.067 m are considered. The number of wave components is $N$ = 300.

Fig. 13 shows the time histories of the target incident wave and the measured wave elevations. The wave elevations of the incident waves are in agreement with the theoretical results which demonstrates that the irregular wave can be well generated and absorbed by the proposed method. Fig. 14 shows the time histories of the motion of the plunger. It can be found that the plunger with gap requires a greater motion response than the plunger without gap to obtain the target incident wave.

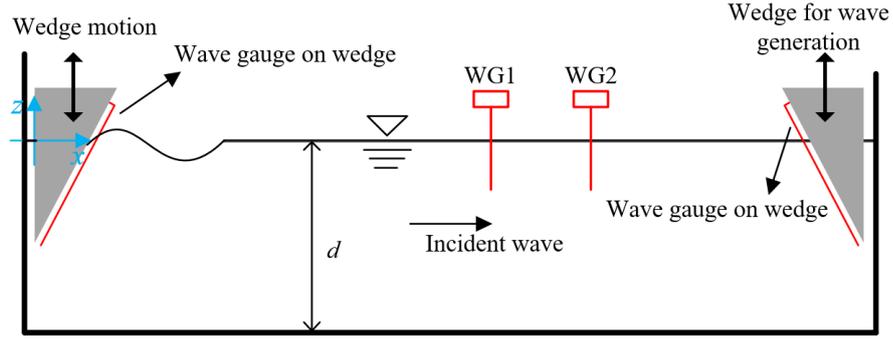

Fig. 12 Schematics of the 2D numerical wave flume

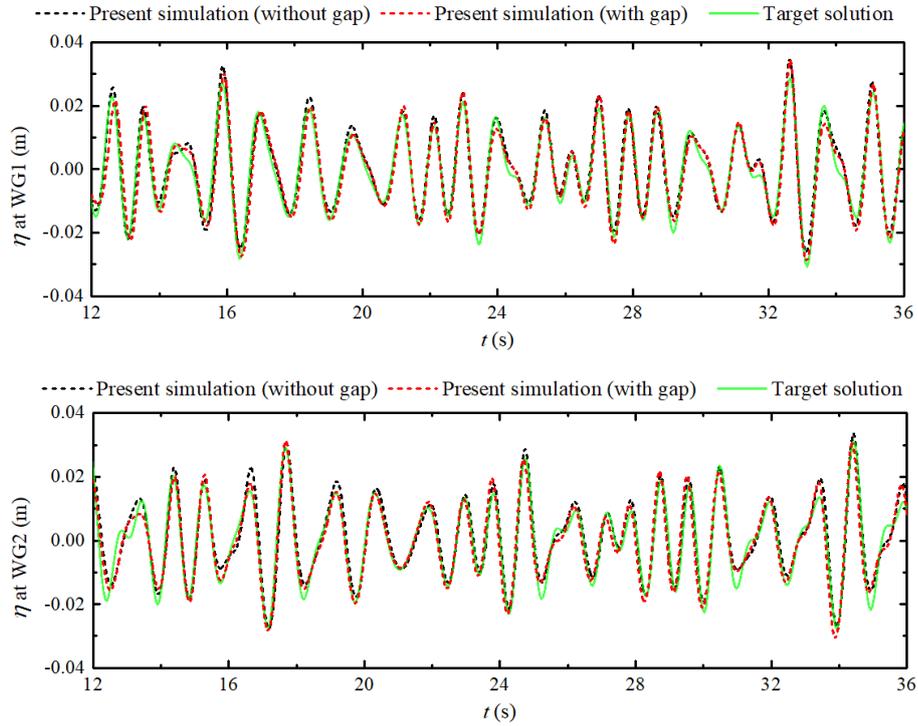

Fig. 13 Time histories of the target incident wave and the measured wave elevations.

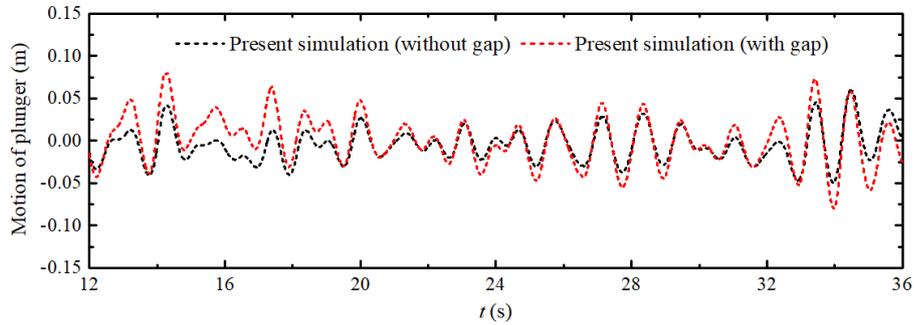

Fig. 14 Time histories of the motion of plunger.

**5.4 Validation of the learned ANN on absorbing wave-making**

In this subsection, we investigate the performance of the wedge-shaped plunger wave-making system in eliminating the re-reflection waves. Similar to that in subsection 4.3, a standing wave is expected to be generated in the computational domain. The wavemaker is located at the one end while the wall is 8*L* away from the wavemaker. The targeting incident wave input to the wavemaker is a second-order Stokes wave. The incident wave period is 0.9 s and its amplitude is set as 0.015 m. Two wave gauges WG1 and WG2 are employed to record the wave evolution, located at antinode (equal to *L*/2 from the right-side wall) and at node (equal to *L*/4 from the right-side wall), respectively. VG is the location ($z = -0.06$ m) where numerical orbital velocities are analyzed.

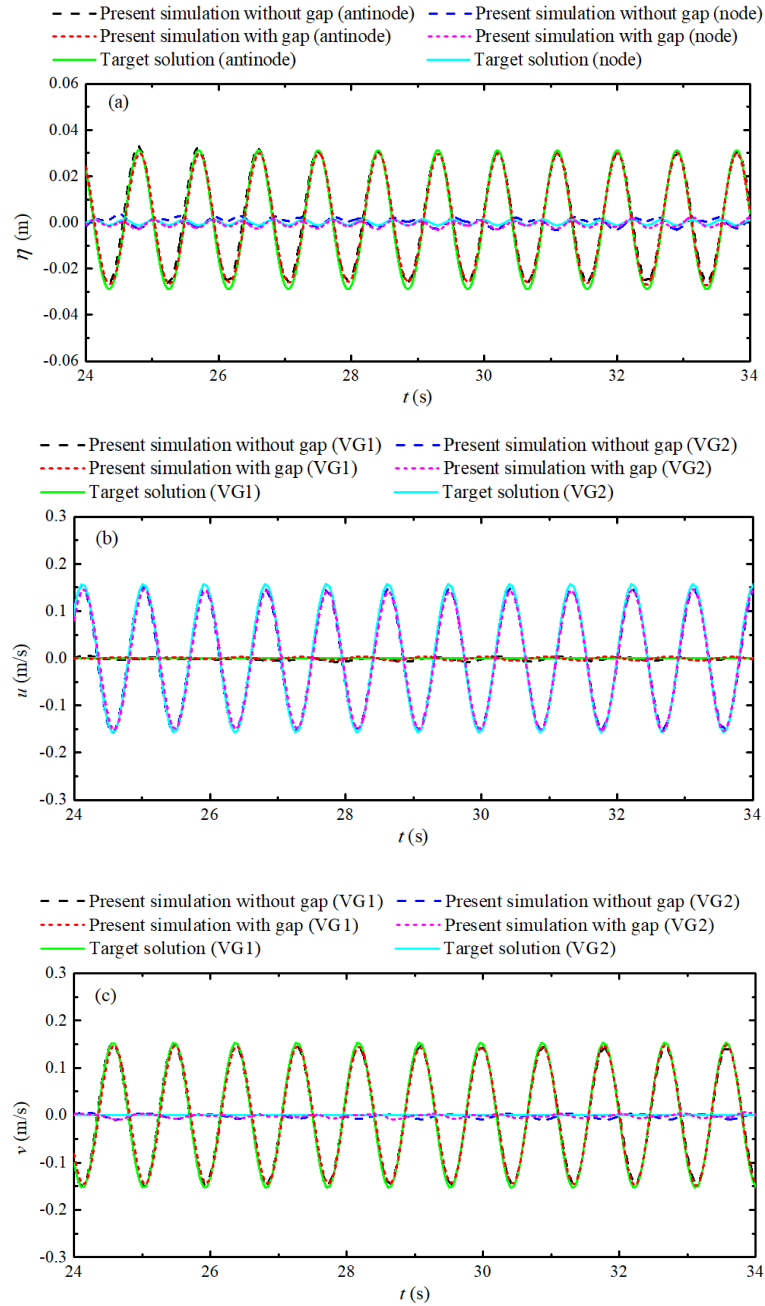

Fig. 16 Time series at node and antinode for the orbital velocity and the free-surface elevation.

Fig. 16 shows the time series at node and antinode, including the free-surface elevation, the horizontal and vertical components of the orbital velocity. The target solutions are also plotted for comparison, assuming perfect reflection at the right-side wall. As shown in Fig. 16(a), the amplitudes at node are very small. Though the amplitudes at antinode are slightly smaller than that of the target solution, the ability of the proposed wave-making system in eliminating the re-reflection waves is gratifying.

## 6. Case III: Cylinder-shaped plunger absorbing wavemaker

Taking a cylinder-shaped plunger as the third example. Data generation will be given in subsection 6.1. Then, the verification of cylinder-induced solitary wave will be conducted in subsection 6.2, while the verification of cylinder-induced New-year wave will be performed in subsection 6.3.

### 6.1 Data generation

As mentioned in the introduction, the advantages of this wave absorption system are two-fold. Firstly, a wide variety of wave conditions with arbitrary shape wavemakers can be applied quickly. Secondly, the transfer function required to absorb reflecting waves can also be used to generate a specified wave. To illustrate these advantages, we use two cylinder plungers with different shapes to generate solitary wave and New-year wave respectively.

The diagram of the 2D numerical wave flume with plunger A is shown in Fig. 17(a). The still water depth is $d = 0.5$ m and the flume length is $12L$, where $L$ is the characteristic wavelength. The initial (zero-position) bottom corner of plunger is located in $z = -D_{plunger} = -0.25$ m. The lower part of the cylinder was an immersed semi-circle with a radius $r = 2$ m. The superstructure was a rectangle, which had a width of $B = 0.6$ m. Wave damping zones are placed on the right sides of the flume with a width of $6L$. The wave gauge mounted in front of the wavemaker is employed to record the free-surface elevation. The diagram of the 2D numerical wave flume with plunger B is shown in Fig. 17(b). The initial (zero-position) bottom corner of plunger is located in $z = -D_{plunger} = -0.2$ m. The cylinder is an immersed semi-circle with a radius $r = 1.5$ m.

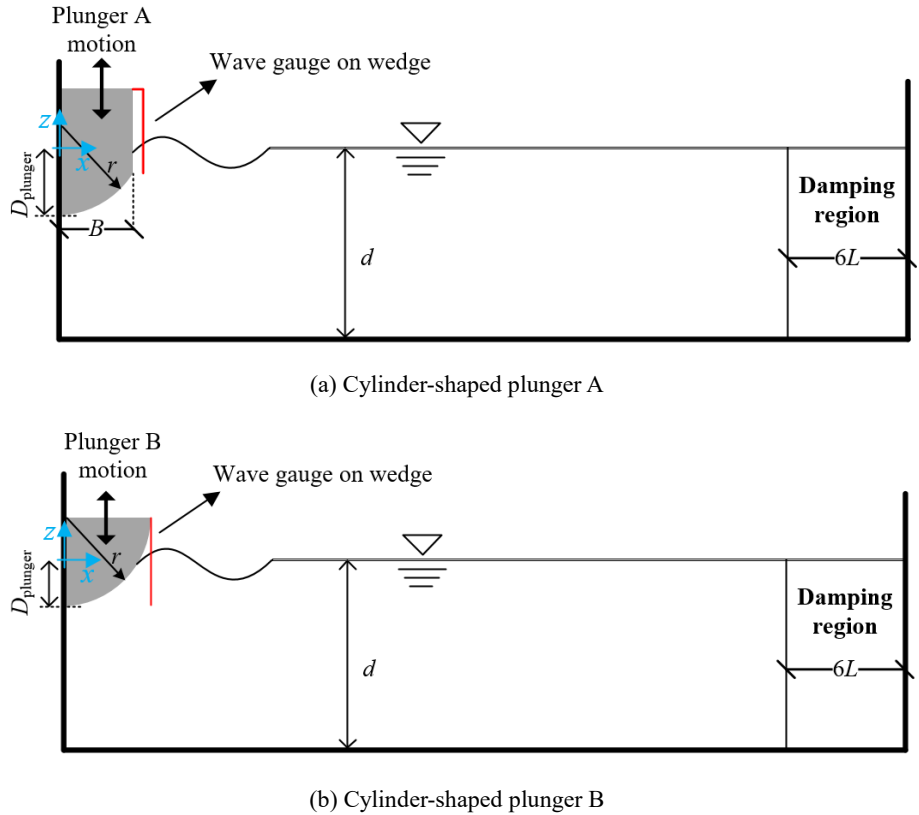

(a) Cylinder-shaped plunger A

(b) Cylinder-shaped plunger B

Fig. 17 Schematics of the 2D numerical wave flume

Similar to section 4.1, we force the sinusoidal motion of the wavemaker to generate a series of regular waves to generate sufficient data for the training of the neural network models. The motion parameters for plungerA-induced waves are listed in Table 6, while those for plungerB-induced waves are listed in Table 7. Each case only needs to be run about three periods of the sinusoidal motion of the wavemaker.

Table 6 The sinusoidal motion parameters of plungerA-induced waves for data generation

|  | T=0.75 s | T=0.95 s | T=1.1 s | T=1.4 s |
|---|---|---|---|---|
| Stroke=0.02 m | Case1 | Case2 |  |  |
| Stroke=0.04 m | Case3 | Case4 | Case5 |  |
| Stroke=0.06 m |  | Case6 | Case7 |  |
| Stroke=0.09 m |  | Case8 | Case9 | Case10 |

Table 7 The sinusoidal motion parameters of plungerB-induced waves for data generation

|  | T=0.8 s | T=0.95 s | T=1.1 s | T=1.4 s | T=2.0 s | T=4.0 s | T=9.0 s |
|---|---|---|---|---|---|---|---|
| Stroke=0.02 m | Case1 | Case2 | Case3 | Case4 |  |  |  |
| Stroke=0.04 m |  | Case5 | Case6 | Case7 |  |  |  |
| Stroke=0.10 m |  |  | Case8 |  | Case9 | Case10 |  |
| Stroke=0.20 m |  |  |  |  | Case11 | Case12 |  |
| Stroke=0.38 m |  |  |  |  |  | Case13 | Case14 |

**6.2 Validation of the learned ANN on generating solitary wave**

In this part, we aim to generate solitary waves by the proposed ANN wave-making system. Although using piston-type wavemakers to generate solitary waves is common, there is hardly any research on the universal method applied to wave flumes equipped with plunger-type wavemakers. Due to the complex geometry, the research for theoretical methods on the plunger-type solitary wave generation is difficult. Recently, He et al. (2021) derived new formulae for the precise descent of plunger-type wavemakers and specified its constraints on the produced wave height are given. However, once the shape of the plunger is changed or an unusual shape is given, the formulae must be deduced again or even has no solution. Significantly, the ANN wave-making system proposed in this paper can be used to control the plunger to produce target solitary waves very simply.

Here, we take plunger B as the example to generate solitary waves. The shape parameters of plunger B are identical to those in subsection 6.1, except changing the $D_{\text{plunger}}$ to = 0.005 m. The still water depth is $d = 0.5$ m and the flume length is 27 m long. Five groups of pressure-velocity probes PV1− PV5 are mounted at $x = 6.0$ m, 6.4 m, 7.0 m, 7.6 m and 8.0 m, respectively. It should be noted that the target solitary wave (input wave profile data) is derived from the third-order Grimshaw solitary wave solution (Grimshaw, 1970; Grimshaw, 1971):

$$\eta(x,t) = d\left[\varepsilon s^2 - \frac{3}{4}\varepsilon^2 s^2 q^2 + \varepsilon^3 \left(\frac{5}{8}s^2 q^2 - \frac{101}{80}s^4 q^2\right)\right] \quad (10)$$

where $\varepsilon = H/d$,

$$s = \text{sech}\left(\frac{\alpha(x-Ct)}{d}\right) \quad (11)$$

$$q = \tanh\left(\frac{\alpha(x-Ct)}{d}\right) \quad (12)$$

and the coefficient $\alpha$ is:

$$\alpha = \sqrt{\frac{3}{4}\varepsilon}\left(1 - \frac{5}{8}\varepsilon + \frac{71}{128}\varepsilon^2\right) \quad (13)$$

$$C^2 = gd\left(1 + \varepsilon - \frac{1}{20}\varepsilon^2 - \frac{3}{70}\varepsilon^3\right) \quad (14)$$

In this study, for small amplitude solitary waves $\varepsilon < 0.2$, the 2rd-order analytical velocity distribution is used:

$$\frac{u}{\sqrt{gd}} = \frac{H}{d}\left\{1+\frac{H}{d}\left[1-\frac{3}{2}\left(\frac{z}{d}\right)^2\right]\right\}\operatorname{sech}^2(\varphi) - \frac{1}{4}\left(\frac{H}{d}\right)^2\left[7-9\left(\frac{z}{d}\right)^2\right]\operatorname{sech}^4(\varphi) \quad (15)$$

$$\frac{v}{\sqrt{gd}} = \sqrt{3}\frac{z}{d}\left(\frac{H}{d}\right)^{3/2}\left\{1+\frac{H}{d}\left[1-\frac{1}{2}\left(\frac{z}{d}\right)^2\right]\right\}\operatorname{sech}^2(\varphi)\tanh(\varphi) - \frac{\sqrt{3}}{2}\frac{z}{d}\left(\frac{H}{d}\right)^{5/2}\left[7-3\left(\frac{z}{d}\right)^2\right]\operatorname{sech}^4(\varphi)\tanh(\varphi) \quad (16)$$

where $\varphi = 2\pi(x/\lambda - t/T) + \pi$. For solitary waves $\varepsilon > 0.2$, the 3rd-order analytical velocity distribution is used:

$$\frac{u}{\sqrt{gd}} = \varepsilon s^2 + \varepsilon^2\left\{-\frac{3}{4}s^2 + s^2 q^2 + \left(\frac{z}{d}\right)^2\left(\frac{3}{4}s^2 - \frac{9}{4}s^2 q^2\right)\right\}$$
$$+\varepsilon^3\left\{\frac{21}{40}s^2 - s^2 q^2 - \frac{6}{5}s^4 q^2 + \left(\frac{z}{d}\right)^2\left(-\frac{9}{4}s^2 + \frac{15}{4}s^2 q^2 + \frac{15}{2}s^4 q^2\right) + \left(\frac{z}{d}\right)^4\left(\frac{3}{8}s^2 - \frac{45}{16}s^4 q^2\right)\right\} \quad (17)$$

$$\frac{v}{\sqrt{gd}} = \sqrt{3\varepsilon}\left(\frac{z}{d}\right)s^2 q\left[\varepsilon + \varepsilon^2\left\{-\frac{3}{8} - 2s^2 + \left(\frac{z}{d}\right)^2\left(-\frac{1}{2} + \frac{3}{2}s^2\right)\right\}\right]$$
$$+\sqrt{3\varepsilon}\left(\frac{z}{d}\right)s^2 q\left[\varepsilon^3\left\{-\frac{49}{640} - \frac{17}{20}s^2 - \frac{18}{5}s^4 + \left(\frac{z}{d}\right)^2\left(-\frac{13}{16} - \frac{25}{16}s^2 + \frac{15}{2}s^4\right) + \left(\frac{z}{d}\right)^4\left(-\frac{3}{40} + \frac{9}{8}s^2 - \frac{27}{16}s^4\right)\right\}\right] \quad (18)$$

Firstly, solitary wave generation with wave height $H=0.04$ m is tested. The motion of the plunger is given in Fig. 18, while snapshots of the free-surface elevation and velocity field are displayed in Fig. 19. Water surges in front of the cylinder and propagates downstream with the acceleration motion of the cylinder. Then, the cylinder slows down after the wave crest is formed, which lowers the free-surface elevation in front of the cylinder. Though the main pulse is followed by redundant trailing waves, these trailing waves are sufficiently small to be ignored. Finally, a complete solitary wave is gradually produced.

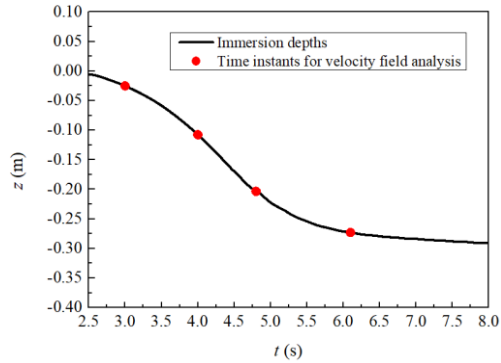

Fig. 18 Time series of the motion of the plunger

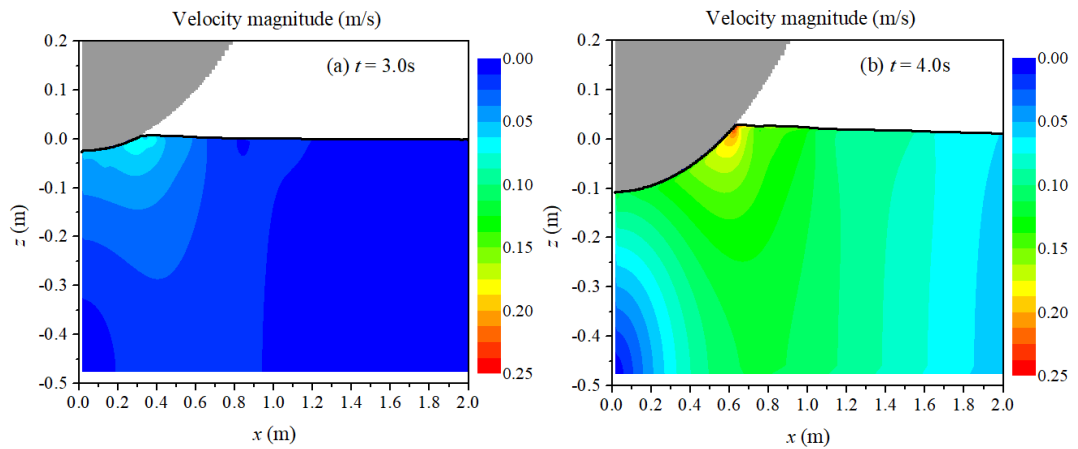

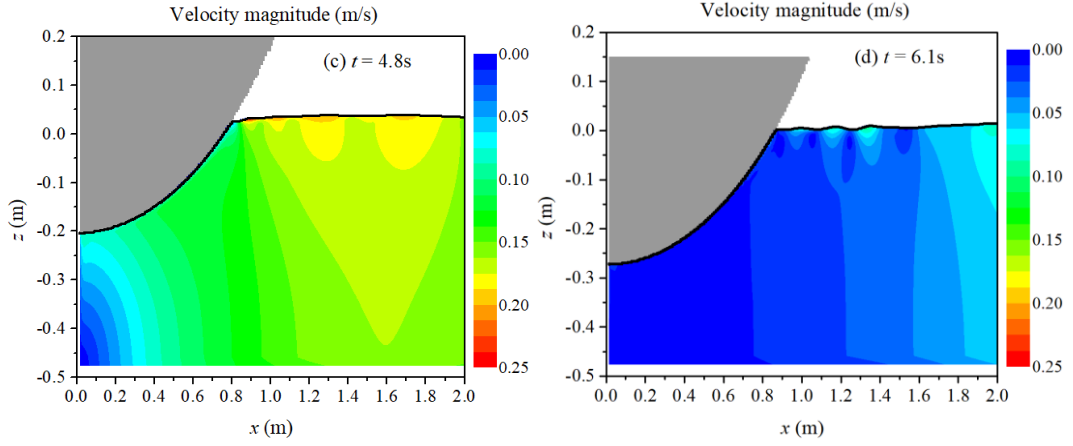

Fig. 19 Evolution of velocity fields during the cylinder descent in water.

The wave profiles and velocity fields are examined by comparing present simulation results with the 2rd-order analytical solutions, shown in Fig. 20. The agreement on water particle velocity is satisfactory. One could notice differences between present simulation results and theoretical results that are depicted in Fig. 20. Interestingly, a hump initially rides on the wave back, which presents a similar phenomenon observed in He et al. (2021). As the same phenomenon occurs with completely different wave-making methods, it may eliminate the defect of the wave-making method. A possible explanation for these differences is the unsteady wave celerity caused by numerical dissipation or the theory assumption that the waveform is permanent. However, the hump gradually integrates into the trailing waves, which has little effect on the quality of the solitary wave.

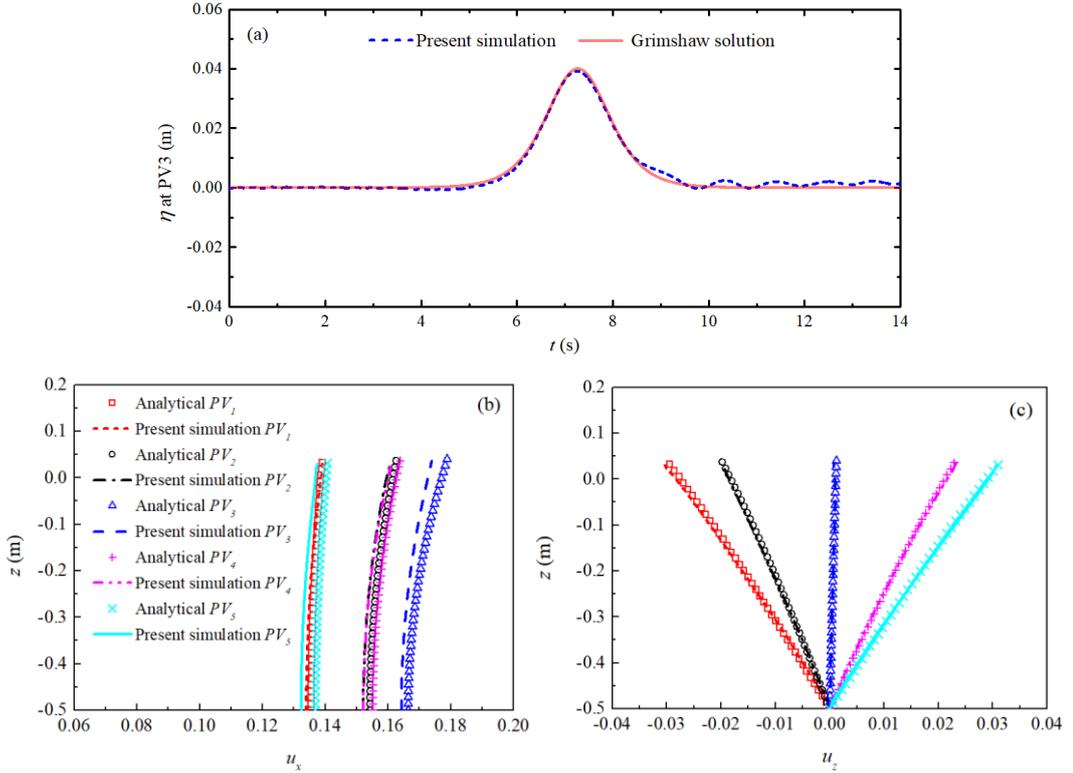

Fig. 20 Comparisons of the computed and analytical wave profiles and water particle velocities (wave height $H$=0.04 m).

Then, solitary wave generation with wave height $H$=0.12 m is tested. The wave profiles and velocity fields are examined by comparing present simulation results with the 3rd-order analytical solutions, shown in Fig. 21. Good agreements on wave profiles and water particle velocity can also be confirmed. The only difference is that the computed horizontal velocity is a little smaller than the analytical solution, and the discrepancy increases with the increase of depth, which presents a similar phenomenon observed in He et al. (2021). Whatever the reason, the reliability of our wave-making method on solitary wave generation is verified.

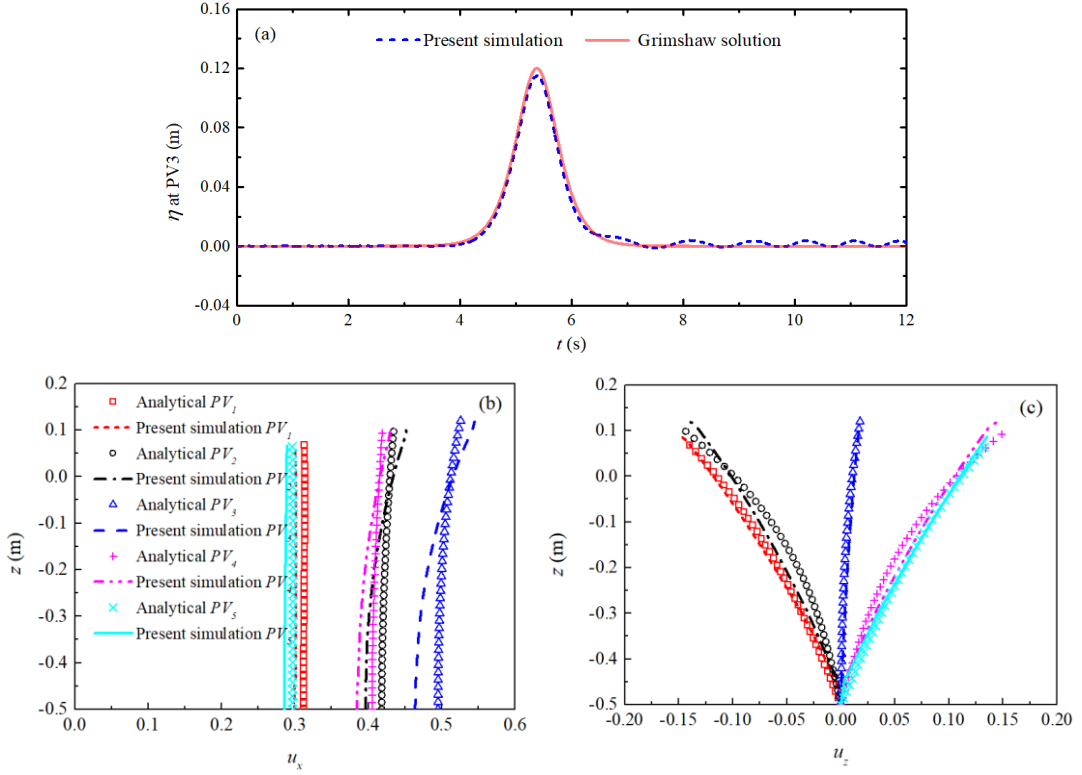

Fig. 21 Comparisons of the computed and analytical wave profiles and water particle velocities (wave height $H$=0.12 m).

**6.3 Validation of the learned ANN on generating New-year wave**

To accurately reproduce a given surface elevation time trace, wave flumes always are required to control wavemaker action to generate desired water waves. In this part, we aim to generate the New-year wave (Haver and Anderson, 2000) by the proposed ANN wave-making system. The wave profile data input to the wavemaker is extracted from the experimental data obtained by Clauss et al. (2011). Here, we take plunger A as the example to generate New-year waves. The shape parameters of plunger A are identical to those in subsection 6.1. The experimental model scale used in Clauss et al. (2011) is 1 : 81, while the numerical model scale is defined as 1 : 243. The still water depth is $d$ = 0.5 m and the flume length is 16 m long. The location of the wave gauge on the plunger in the present model corresponds to the location of $x$ = 21.3 m in the experiment model where the free-surface elevation time trace data is available (so data can be used as input). Two wave gauges WG1 (which is located at $x$ = 2.217 m in the present model) and WG2 (which is located at $x$ = 2.55 m in the present model) correspond to the location of $x$ = 27.95 m and $x$ = 28.95 m in the experiment model. It should be noted that the record New-year wave would be reproduced at WG2.

Fig. 22 shows the time histories of the targeting incident wave and the measured wave elevations. The wave elevations of the incident waves are in agreement with the reference ones (Clauss et al., 2011), which demonstrates that the given surface elevation time trace can be accurately reproduced by the proposed wave-making system. Based on the presented analysis of the proposed wave-making system, it can be concluded that the transfer function mapped by ANN is a simple and efficient tool for desired wave generation.

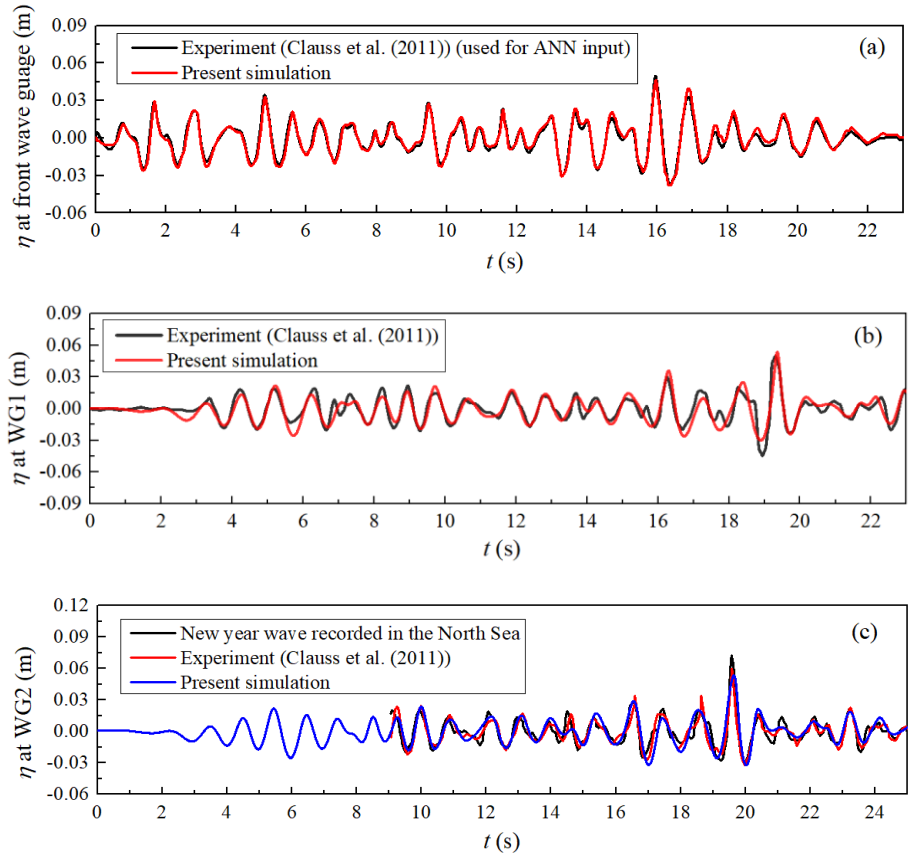

Fig. 22 Time histories of the desired incident wave and the measured wave elevations

## 7. Conclusion

In this paper, a universal framework is proposed for wavemakers with a front-mounted wave gauge to generate and absorb waves. The neural networks are trained to establish the transfer function between the free-surface elevation on the wavemaker and the wavemaker velocity. Once the target wave profiles in front of the wavemaker are given, it can realize generating waves and absorbing reflected waves at the same time. The penalty term is used to embed the physical feature into the loss function and prevent neural networks from overfitting. Taking piston and plunger wavemakers as examples, both the wave generation and absorption are simulated by an in-house numerical solver. In addition, solitary waves and New-year waves are generated to validate the framework. The main conclusions are summarized as follows:

By introducing data augmentation to enrich the input variables, the convergence error is smaller and the error distribution is more concentrated near zero error. From the perspective of wave-making mechanisms, the variable of velocities of the free-surface elevation increases the momentum information, while the couple of free-surface elevation in front of the wavemaker and the position of the wavemaker provide the information of waterplane area.

For regular waves full reflection case, it is depicted that the proposed framework could effectively absorb re-reflection waves and it is possible to produce pure standing waves. In the case of New-year waves and solitary waves, it is possible to accurately generate the desired wave-elevation time series, given the target wave profiles in front of the wavemaker.

This research demonstrates the potential for the use of machine learning technology in generating and absorbing waves. Further investigation and development should be done to apply this method to practical experiments and improve the utilization of the technology.

**Acknowledgment**

This study was partially supported by the National Natural Science Foundation of China (GrantNos. 51979245, 51679212).

**Declaration of interests**

The authors report no conflict of interest.

**Reference**

Calderer, A., Kang, S., Sotiropoulos, F., 2014. Level set immersed boundary method for coupled simulation of air/water interaction with complex floating structures. Journal of Computational Physics 277, 201–227. https://doi.org/10.1016/j.jcp.2014.08.010

Chen, L.-W., Cakal, B.A., Hu, X., Thuerey, N., 2021. Numerical investigation of minimum drag profiles in laminar flow using deep learning surrogates. J. Fluid Mech. 919, A34. https://doi.org/10.1017/jfm.2021.398

Chorin, A.J., 1968. Numerical solution of the Navier-Stokes equations. Mathematics of Computation 22, 745–745. https://doi.org/10.1090/S0025-5718-1968-0242392-2

Christensen, M., Frigaard, P., 1994. Design of absorbing wave maker based on digital filters, in: Proceedings of International Symposium: Waves-Physical and Numerical Modelling. Vancouver, pp. 100–109.

Clauss, G.F., Klein, M., 2011. The new year wave in a seakeeping basin: generation, propagation, kinematics and dynamics. Ocean Engineering 38, 1624–1639. https://doi.org/10.1016/j.oceaneng.2011.07.022

Eldrup, M.R., Andersen, T.L., 2019. Applicability of nonlinear wavemaker theory. Journal of Marine Science and Engineering 7.

Ellix, D., Arumugam, K., 1984. An experimental study of waves generated by an oscillating wedge. Journal of Hydraulic Research 22, 299–313. https://doi.org/10.1080/00221688409499367

Grimshaw, R., 1971. The solitary wave in water of variable depth. Part 2. Journal of Fluid Mechanics 46, 611–622. https://doi.org/10.1017/S0022112071000739

Grimshaw, R., 1970. The solitary wave in water of variable depth. Journal of Fluid Mechanics 42, 639–656. https://doi.org/10.1017/S0022112070001520

Haver, S., Andersen, O., 2000. Freak waves: Rare realizations of a typical population or typical realizations of a rare population?, in: Proceedings of the International Offshore and Polar Engineering Conference. Seattle, USA, pp. 123–130.

He, M., Khayyer, A., Gao, X., Xu, W., Liu, B., 2021. Theoretical method for generating solitary waves using plunger-type wavemakers and its Smoothed Particle Hydrodynamics validation. Applied Ocean Research 106, 102414. https://doi.org/10.1016/j.apor.2020.102414

Hicks, J.B.H., Bingham, H.B., Read, R.W., Engsig-Karup, A.P., 2021. Nonlinear wave generation using a heaving wedge. Applied Ocean Research 108, 102540. https://doi.org/10.1016/j.apor.2021.102540

Hirakuchi, H., Kajima, R., Kawaguchi, T., 1990. Application of a piston-type absorbing wavemaker to irregular wave experiments. Coastal Engineering in Japan 33, 11–24. https://doi.org/10.1080/05785634.1990.11924520

Hornik, K., 1991. Approximation capabilities of multilayer feedforward networks. Neural Networks 4, 251–257.

Johannes, Spinneken, and, Chris, Swan, 2012. The operation of a 3D wave basin in force control. Ocean Engineering 55, 88–100.

Khait, A., Shemer, L., 2019. Nonlinear wave generation by a wavemaker in deep to intermediate water depth. Ocean

Engineering 182, 222–234.

Kutz, J.N., 2017. Deep learning in fluid dynamics. J. Fluid Mech. 814, 1–4. https://doi.org/10.1017/jfm.2016.803

Lee, S., You, D., 2019. Data-driven prediction of unsteady flow over a circular cylinder using deep learning. J. Fluid Mech. 879, 217–254. https://doi.org/10.1017/jfm.2019.700

Mahjouri, S., Shabani, R., Rezazadeh, G., Badiei, P., 2020. Active control of a piston-type absorbing wavemaker with fully reflective structure. China Ocean Engineering 34, 730–737. https://doi.org/10.1007/s13344-020-0066-9

Mello, P.D., Carneiro, M.L., Tannuri, E.A., Kassab, F., Marques, R.P., Adamowski, J.C., Nishimoto, K., 2013. A control and automation system for wave basins. Mechatronics 23, 94–107.

Nikseresht, A.H., Bingham, H.B., 2020. A numerical investigation of gap and shape effects on a 2d plunger-type wave maker. Journal of Marine Science and Application 19, 101–115. https://doi.org/10.1007/s11804-020-00135-5

Osher, S., Fedkiw, R., Piechor, K., 2004. Level set methods and dynamic implicit surfaces. Applied Mechanics Reviews 57, B15–B15. https://doi.org/10.1115/1.1760520

Osher, S., Fedkiw, R.P., 2001. Level set methods: an overview and some recent results. Journal of Computational Physics 169, 463–502. https://doi.org/10.1006/jcph.2000.6636

Peskin, C.S., 2002. The immersed boundary method, in: Iserles, A. (Ed.), Acta Numerica 2002. Cambridge University Press, pp. 479–518. https://doi.org/10.1017/CBO9780511550140.007

Schäffer, H.A., Jakobsen, K., 2003. Non-linear wave generation and active absorption in wave flumes. Presented at the Proceedings of Long Waves Symposium 2003 in Parallel with XXX IAHR Congress, Greece, pp. 69–77.

Schäffer, H.A., Klopman, G., 2000. Review of multidirectional active wave absorption methods. Journal of Waterway, Port, Coastal, and Ocean Engineering 126, 88–97. https://doi.org/10.1061/(ASCE)0733-950X(2000)126:2(88)

Schmitt, P., 2017. Steps towards a self calibrating, low reflection numerical wave maker using narx neural networks. Presented at the International Conference on Computational Methods in Marine Engineering.

Schmitt, P., Gillan, C., Finnegan, C., 2021. On the use of artificial intelligence to define tank transfer functions (preprint). ENGINEERING. https://doi.org/10.20944/preprints202110.0252.v1

Seaïd, M., 2002. Semi-lagrangian integration schemes for viscous incompressible flows. Computational Methods in Applied Mathematics 2, 392–409. https://doi.org/10.2478/cmam-2002-0022

Spinneken, J., Swan, C., 2009. Second-order wave maker theory using force-feedback control. Part I: A new theory for regular wave generation. Ocean Engineering 36, 539–548.

Srivastava, N., Hinton, G., Krizhevsky, A., Sutskever, I., Salakhutdinov, R., 2014. Dropout: a simple way to prevent neural networks from overfitting. Journal of Machine Learning Research 15, 1929–1958.

Timmerberg, S., Börner, T., Shakeri, M., Ghorbani, R., Alam, M.-R., 2015. The "Wave Bridge" for bypassing oceanic wave momentum. Journal of Ocean Engineering and Marine Energy 1, 395–404. https://doi.org/10.1007/s40722-015-0028-0

van Leer, B., 1979. Towards the ultimate conservative difference scheme. V. A second-order sequel to Godunov's method. Journal of Computational Physics 32, 101–136. https://doi.org/10.1016/0021-9991(79)90145-1

Wu, Y.-C., 1991. Waves generated by a plunger-type wavemaker. Journal of Hydraulic Research 29, 851–860. https://doi.org/10.1080/00221689109498963

Wu, Y.-C., 1988. Plunger-type wavemaker theory. Journal of Hydraulic Research 26, 483–491. https://doi.org/10.1080/00221688809499206

Xie, Y., Zhao, X., 2021. Sloshing suppression with active controlled baffles through deep reinforcement learning–expert demonstrations–behavior cloning process. Physics of Fluids 33, 017115. https://doi.org/10.1063/5.0037334

Xie, Y., Zhao, X., Luo, M., 2022. An active-controlled heaving plate breakwater trained by an intelligent framework based on deep reinforcement learning. Ocean Engineering 244, 110357. https://doi.org/10.1016/j.oceaneng.2021.110357

Yabe, T., Xiao, F., Utsumi, T., 2001. The constrained interpolation profile method for multiphase analysis. Journal of Computational Physics 169, 556–593. https://doi.org/10.1006/jcph.2000.6625

Yang, H., Li, M., Liu, S., Chen, F., 2016. An iterative re-weighted least-squares algorithm for the design of active abso- rbing wavemaker controller. Journal of Hydrodynamics, Ser. B 28, 206–218. https://doi.org/10.1016/S1001-6058(16)60622-4

Zhang, N., Zheng, Z.C., 2007. An improved direct-forcing immersed-boundary method for finite difference applications. Journal of Computational Physics 221, 250–268. https://doi.org/10.1016/j.jcp.2006.06.012

Zhao, X., Hu, C., 2012. Numerical and experimental study on a 2-D floating body under extreme wave conditions. Applied Ocean Research 35, 1–13. https://doi.org/10.1016/j.apor.2012.01.001

Zhao, X., Ye, Z., Fu, Y., Cao, F., 2014. A CIP-based numerical simulation of freak wave impact on a floating body. Ocean Engineering 87, 50–63. https://doi.org/10.1016/j.oceaneng.2014.05.009